\documentclass{article}

\usepackage{iclr2027/iclr2027_conference,times}
\iclrfinalcopy

\usepackage[utf8]{inputenc}
\usepackage[T1]{fontenc}
\usepackage{url}
\usepackage{booktabs}
\usepackage{amsfonts}
\usepackage{amsmath}
\usepackage{nicefrac}
\usepackage{microtype}
\usepackage{xcolor}
\usepackage{array}
\usepackage{float}
\usepackage{longtable}
\usepackage{fvextra}
\usepackage{multirow}
\usepackage{tikz}
\usetikzlibrary{arrows.meta,positioning,shapes.geometric}
\usepackage{hyperref}

\definecolor{linkblue}{RGB}{32,77,120}
\hypersetup{
  colorlinks=true,hypertexnames=false,
  linkcolor=linkblue,citecolor=linkblue,urlcolor=linkblue,
  pdftitle={Beyond LLM Judge: Mechanical Verification for Agentic Autoformalization},
  pdfauthor={Anna Mei and Michael R. Douglas}
}
\newcolumntype{L}[1]{>{\raggedright\arraybackslash}p{#1}}

\title{Beyond LLM Judge: Mechanical Verification for Agentic Autoformalization}

\author{Anna Mei\\Harvard University\\\texttt{amei@g.harvard.edu}
\And Michael R. Douglas\\CMSA, Harvard University and AletheAI, Inc\\\texttt{mdouglas@cmsa.fas.harvard.edu}}
\date{}

\begin{document}

\maketitle
\fancyhead{}

\begin{abstract}
We introduce a novel method for verifying Lean formalizations that goes beyond LLM-judged faithfulness. Unlike most verification methods, which rely on LLMs or trained models, our method combines LLM-based semantic review with mechanical numerical and vacuity checks, providing checkable feedback and detecting errors missed by LLM judges. We apply this method, along with our Extract-Verify-Proof autoformalization pipeline, to construct the Formalized Library of Mathematical Functions (FLMF), a partial formalization of the NIST Digital Library of Mathematical Functions (DLMF). On three labeled datasets, our method achieves the lowest accepted-error rate among the evaluated methods, measured as the fraction of incorrect statements accepted by the verification method. Expert review of the results also identifies translation and source defects in existing datasets. Code available at \url{https://github.com/annamei1/FLMF-Ch7ErrFunc-Release.git}.
\end{abstract}

\section{Introduction}

With the rapid development of agentic coding assistants, the formalization of mathematics has become very accessible to researchers and has received a lot of attention. However, to this day, the most trusted and widely used library is still Mathlib \citep{mathlib}, whose correctness is guaranteed by careful human review and which is therefore trusted by its downstream users. For the same reason, it also faces a slow review process and limited coverage of existing mathematics, blocking the formalization of more advanced mathematical work. Efforts have been made to resolve this problem by building agentic systems to formalize textbooks, research blueprints, and dependency-guided libraries at increasing scale \citep{gloeckle2026,rammal2026,wang2026,cabral2026,milikic2026,douglas2026}. 

In contrast to Mathlib, autoformalizations of this kind face severe challenges to their reliability. An autoformalizer can change a domain, index, sign, or branch convention, or introduce contradictory hypotheses, and still produce a provable statement. Audits of formal benchmarks document such defects \citep{an2026}.

We therefore propose a verification method that combines numerical verification, an LLM check, and a vacuity check, along with an agentic autoformalization pipeline, to allow the creation of trustworthy reusable formal libraries. We chose the Digital Library of Mathematical Functions (DLMF) as our testing ground, and propose the Formalized Library of Mathematical Functions (FLMF) as the product of our pipeline and verification method.

This paper makes three contributions:
\begin{itemize}
\item \textbf{Numerical and vacuity verification method} (Section~\ref{sec:verification}). We introduce numerical and vacuity checks on Lean statements, which are mechanical and provide checkable feedback on points of failure. As indicated by their names, a numerical check verifies numerical validity of a Lean statement, and a vacuity check looks for vacuous formalization, for example a tautological definition. So far, all verification in autoformalization relies on LLM judgements, and our method complements the ambiguity and arbitrariness of an LLM judge's verdicts. Its feedback is tractable and informative, and passing these checks adds another layer of certainty about the correctness of a Lean statement.
\item \textbf{Autoformalization pipeline} (Section~\ref{sec:pipeline}). Our pipeline divides the formalization process into three stages: extract, verify, and prove. We emphasize human gating at each stage, requiring every item either to pass a gate or to be documented with a reason for failure. We then use an auditor agent to revise those documents against written guidelines. We also keep track of all statement corrections to allow future studies of them.
\item \textbf{Corrections to existing datasets} (Section~\ref{sec:experiments}). Expert review of our verification flags identifies multiple defects in existing datasets. Among the statements that are labeled correct, we found 2 defects among ProofNet items labeled faithful by Ground False, 61 defects among correct-labeled Beyond Compilation items, and 205 defects in RLM25 (Table~\ref{tab:flag-review}). We will release corrected statements for confirmed defects. 
\end{itemize}

\section{Related Work}
\label{sec:related}

\paragraph{Evaluating formalization faithfulness.}
Some recent works use reference-based metrics to evaluate the faithfulness of formalization: BEq and its variants check bidirectional provable equivalence \citep{liu2025rethinking,poiroux2025reliable}, GTED compares operator trees \citep{liu2025gted}, and ASSESS combines semantic and structural similarity \citep{liu2026assess}. These metrics need a reference formal statement, which is a limitation when it comes to the formalization of natural language statements. On the other hand, reference-free methods score the informal--formal pair directly. FormalAlign trains a model to predict alignment from sequence certainty and representation similarity \citep{lu2025formalalign}. LeanScorer breaks a statement into components and has an LLM judge each one \citep{xuejun2025mathesis}. Beyond Compilation evaluates LLM judges for faithful natural-language-to-Lean translation \citep{zhang2026beyond}. FormalMATH uses back-translation followed by LLM judgments to filter formalizations that do not preserve the original problem \citep{yu2025formalmath}.  As a result, Poiroux et al.\ show that automatic metrics disagree with humans on research-level statements and release RLM25 \citep{poiroux2025reliable}. Ground False audits formal benchmarks for kernel-checked but wrong targets, classifies each ProofNet statement's relation to its informal source, and releases both that taxonomy and a corrected ProofNet-Verified \citep{an2026}. In this paper, we compare our verification method against them using formal--informal pair datasets.

\paragraph{Agentic formalization.}
Agentic formalization extends single statement translation to the construction of formal libraries. Gloeckle et al.\ coordinate parallel agents on a shared code base to formalize a graduate textbook \citep{gloeckle2026}, while AutoformBot combines dependency-aware scheduling and collaborative version control to build libraries across multiple textbooks \citep{rammal2026}. At the workflow level, LeanFlow studies how proof workflows and toolsets affect the completion and efficiency of formalizing entire papers \citep{milikic2026}; Requena Pozo et al.\ combine iterative proof refinement, library search, and context management in a minimal agent \citep{requena2026}; and OpenGauss provides an open-source Lean autoformalization harness \citep{opengauss2026}. FormalScience incorporates domain-expert feedback into an agentic pipeline for formalizing scientific reasoning \citep{meadows2026formalscience}. These systems motivate our emphasis on statement verification and human gates within a larger formalization workflow. 

\section{Faithfulness Verification}
\label{sec:verification}

Our verification method consists of three components: a semantic (LLM) check, a numerical check, and a vacuity check. The first two evaluate the Lean theorem statement without the proof body. The vacuity check runs after the proof is constructed and determines if the proof was vacuous. No single check establishes faithfulness, and we treat verification as a stack in which each layer has a distinct failure model and any substantive mismatch blocks automatic acceptance. The following discussion is based on our formalization of DLMF, but a large part of it is applicable to other large-scale formalizations of natural language math. For each theorem, we attached a verification tag, which is \texttt{unverified} by default and turns to \texttt{verified} after passing the checks.

\begin{figure}[H]
\centering
\begin{tikzpicture}[
  scale=0.84,
  transform shape,
  font=\scriptsize,
  box/.style={draw, rounded corners, align=center, inner sep=3.5pt, minimum height=7mm, text width=22mm},
  decision/.style={draw, diamond, aspect=2.2, align=center, inner sep=1.5pt},
  arrow/.style={-{Latex[length=1.8mm]}, semithick}
]
\node[box] (surface) at (0,0.8) {Lean statement\\(\texttt{unverified})};
\node[box] (back) at (2.8,1.45) {Semantic check\\(every theorem)};
\node[box] (numeric) at (2.8,0.15) {Numerical check\\(when possible)};
\node[decision] (fidelity) at (5.6,0.8) {Verified?};
\node[box] (verified) at (8.2,0.8) {\texttt{verified} stamp};
\node[box] (review) at (5.6,-0.9) {Agent repair and rerun,\\or human review};
\node[box] (proof) at (8.2,-2.4) {Proof completed};
\node[box] (vacuity) at (5.2,-2.4) {Vacuity check};
\node[box, text width=30mm] (outcome) at (1.9,-2.4) {Clean: retain \texttt{verified}\\Flagged: revert to \texttt{unverified}};
\draw[arrow] (surface) -- (back);
\draw[arrow] (surface) -- (numeric);
\draw[arrow] (back) -- (fidelity);
\draw[arrow] (numeric) -- (fidelity);
\draw[arrow] (fidelity) -- node[above] {pass} (verified);
\draw[arrow] (fidelity) -- node[right] {fail} (review);
\draw[arrow] (verified) -- (proof);
\draw[arrow] (proof) -- (vacuity);
\draw[arrow] (vacuity) -- (outcome);
\end{tikzpicture}
\caption{The verification workflow. A theorem is \texttt{verified} if it receives a semantic \texttt{FAITHFUL} and no numerical or vacuity \texttt{FAIL}.}
\label{fig:verification-flow}
\end{figure}
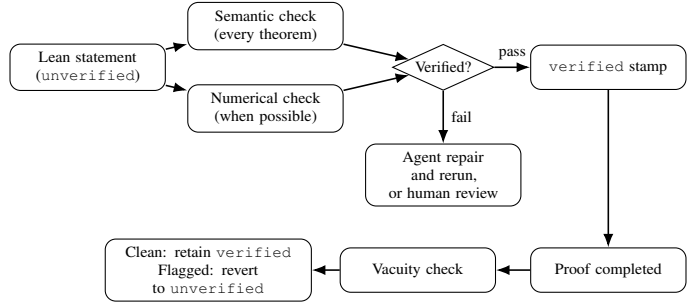

\subsection{Semantic Check} 

The semantic checks consist of a direct LLM inspection and a back-translation check. The direct inspection passes both the Lean statement and the informal statement to an LLM judge, and the judge returns a verdict about the faithfulness of the Lean translation. 

The back-translation check consists of two steps. In the first step, an agent (translator) reads the Lean theorem statement, along with all definitions involved, \textit{without} comments, and translates it into plain English. Masking the comments reduces the risk of hallucination on the agent's side, forcing it to adhere to the exact Lean code. The translation is formatted to include the statement, variables, constraints, and symbols. 

As an example, the translation of \S DLMF5.5.3 $\Gamma(s)\Gamma(1-s)=\pi/\sin(\pi s)$ would be:
\begin{quote}\scriptsize\ttfamily\raggedright
STATEMENT: For any complex number $z$ that is not an integer,
$\Gamma(z)\Gamma(1-z)=\pi/\sin(\pi z)$.

VARIABLES: $z$: A complex number, $z\in\mathbb{C}$.

CONSTRAINTS: $z\neq m$ for every $m\in\mathbb{Z}$.

SYMBOLS: Gamma - the complex Gamma function. Complex.sim - the complex sine function. $\uparrow\pi$ - the mathematical constant $\pi$, cast to a complex number.
\end{quote}

A second agent (judge) reads this English rendering and compares it with the equation extracted from the DLMF source.\footnote{We designed an extractor to pull information from DLMF metadata, which involves an XML file along with the website display, allowing more reliable extraction of equation content and constraints. A sample extracted JSON file is available in Appendix~\ref{app:extraction}.} In addition, we supply a STANDING\_CONSTRAINTS.md document to capture the constraints stated earlier in the source that may be missed by the equation parser.

The direct inspection may be the most intuitive thing to do, but having a back-translation check guarantees that the Lean statement is accurate enough to preserve its meaning after a back-translation. Having two separate agents working independently also avoids the failure mode in which an agent recognizes an inconsistency but chooses to ignore it. The verdicts from the judges fall into three categories: \texttt{FAITHFUL}, \texttt{DIVERGE}, or \texttt{UNCERTAIN}. For the latter two, the judge also documents why it considers the translation unfaithful and waits for further action. When a theorem obtains a \texttt{FAITHFUL} verdict from both the direct inspection and the back-translation check, its verification status can change to \texttt{verified}.

\subsection{Numerical Check} 

When possible\footnote{Exceptions include definitions, limits, infinite sums, etc., that cannot be converted into a numerical check.}, a stated theorem undergoes a numerical check in which a mechanical extractor translates the Lean theorem into evaluations in \emph{mpmath}x, an open-source Python library for arbitrary-precision numerical computation and special functions. A special function is bound to its mpmath counterpart with the appropriate convention. The checker samples points from the stated domain and compares the function values on both sides of the identity. Exact identities use residual gates; asymptotics use ratio or rate gates. Failure of the numerical check vetoes the \texttt{verified} status granted by the back-translation check. 

One natural question is how to guarantee the correctness of the Lean-mpmath binding. This binding is trustworthy because we require that every special function is only defined once, and the binding is attached to the definition. When performing the numerical check, the program traces the definitions used in the Lean statement and uses the corresponding bindings. We can also observe how many times each definition/binding has been used independently for numerical checks, further confirming its reliability.

There are also limitations to the numerical check. As we are working with special functions in DLMF, we can only sample points in the domain and evaluate the functions at those points. A common failure is the inability to identify an incorrect domain for a functional equality. For example, the Lean statement may have a smaller domain than the actual statement, or it may fail to exclude isolated points from the domain. If those points are not selected by the numerical checker, it will fail to flag the error. Therefore, the numerical check should only be used as a partial check for correctness, not a complete faithfulness check. 

\subsection{Vacuity Check} 

A vacuity check scans the theorem statement and proof body to detect empty statements. Simple checks include testing whether trivial proof methods such as \texttt{rfl} or \texttt{simp} close the goal. In addition, we have a checker that scans for goal-independent proofs, which works by replacing the target goal with an opaque conclusion \texttt{VacQ} and checking whether the proof still compiles. If it does, then the proof is goal-independent, and the most likely cause is a set of contradictory hypotheses from which one can derive \texttt{False}\footnote{Any conclusion can be derived from \texttt{False}.}. If the checker flags a vacuous proof, the \texttt{verified} status is also reverted to \texttt{unverified}.

We would like to emphasize that none of these checks is perfect; they all have different domains of effectiveness and limitations, and the stack is undergoing active development as the project progresses. For example, the case that motivated the vacuity check was the original formalization of DLMF 7.13.5: its back-translation was judged faithful, yet proof review found that its completeness and ordering hypotheses for the Fresnel zeros were jointly contradictory. We also made these checks strict on purpose, so we would rather have them flag a correct statement than miss an incorrect one. (Section~\ref{sec:cases} gives more detail on this particular case.)

\section{Autoformalization Pipeline}
\label{sec:pipeline}

The DLMF is organized in chapters and equations. Distinct chapters in the DLMF focus on different types of functions and rarely depend on each other; therefore, parallelization is quite feasible in FLMF. There are also special ``methods'' chapters, which describe mathematical techniques used to establish some of the identities. In these cases, we need to formalize, or at least state, the content of the methods before we can prove the related identities. These dependency relations are captured in a DAG (Directed Acyclic Graph) and are dynamically updated as formalization progresses. When the proving agent attempts to prove a theorem, it first looks it up in the DAG and reports whether infrastructure needed to construct the proof is missing. 

This strategy is not unique to DLMF. Indeed, related works have used similar approaches \citep{wang2026,cabral2026}, and it plays a crucial role in the autoformalization of more sophisticated math objects that have complicated dependencies on each other.

In FLMF, we proceed with formalization in three stages: Extract, Verify, and Prove. 

\begin{table}[H]
\centering
\small
\setlength{\tabcolsep}{4pt}
\begin{tabular}{@{}L{.12\linewidth}L{.39\linewidth}L{.43\linewidth}@{}}
\toprule
Stage & Agent work & Gate on every theorem / equation \tabularnewline
\midrule
DAG & Enumerate equations and dependencies & Open-world gap report\\
Extract & Write typed definitions and theorem statements & (i) Formalized with appropriate tag or (ii) documented in \texttt{DISPOSITIONS.md} \\
Verify & Run numerical and back-translation fidelity checks & (i) Obtained a \texttt{verified} stamp or (ii) documented in \texttt{HUMAN\_REVIEW.md} \\
Prove & Fill frozen theorem bodies in waves & (i) Proved with clean build and 0 \texttt{sorry} or (ii) documented in \texttt{UNPROVEN.md} \\
\bottomrule
\end{tabular}
\caption{Autoformalization pipeline for FLMF.}
\label{tab:pipeline}
\end{table}

\subsection{Extract} 

At this stage, an agent extracts all listed equations in a chapter by their equation numbers. The agent then writes each equation in the DAG and establishes its dependencies on existing nodes. The agent identifies all \emph{definitions} needed to state the extracted equations, for example, the definitions of Bessel functions or theta functions. The definitions are only written once to guarantee consistency in downstream usage. After that, parallel agents are deployed to write Lean statements using these definitions. A complete proof is not required. Each extracted theorem carries 4 tags: DLMF equation number, Fidelity check method, Verification status, and its corresponding DAG node. The Extract workflow is gated on completion: every listed equation in the DLMF has to either (i) appear in FLMF as a theorem with appropriate tags or (ii) be documented in \texttt{DISPOSITIONS.md} with an explanation of why it is not stated. This is an example of a formalized statement:

\begin{quote}\scriptsize\ttfamily\raggedright
/-- **DLMF 5.5.3** --- $\Gamma(s)\Gamma(1-s)=\pi/\sin(\pi s)$ (complex)\\
(DLMF-faithful restriction of `complex\_reflection':$s\neq 0,\pm1,\pm2,\ldots$).\\
DAG: E-5.5.3.\\
Fidelity: numeric PASS\\
Verified: verified\\
-/\\
theorem complex\_reflection\_dlmf\\
\hspace*{1em}(s : $\mathbb{C}$)\\
\hspace*{1em}(\_hs : $\forall$ m : $\mathbb{Z}$, s $\neq$ (m : $\mathbb{C}$)) :\\
\hspace*{1em}Complex.Gamma s * Complex.Gamma (1 - s) =\\
\hspace*{2em}(Real.pi : $\mathbb{C}$) /\\
\hspace*{2em}Complex.sin ((Real.pi : $\mathbb{C}$) * s) := by\\
\hspace*{1em}sorry
\end{quote}

\subsection{Verify} 

When a theorem is first stated, its verification status is \texttt{unverified} by default. A successful verification check, which is the combination of semantic, numerical, and vacuity checks discussed in Section~\ref{sec:verification}, allows a stamper to set this tag to \texttt{verified}. If a theorem fails a verification check, an agent analyzes the reason for failure. For the easy fixes, e.g., missing premises, a wrong sign, or incompatible conventions, an agent fixes the theorem and reruns the verification check; otherwise it is directed to human review. This stage is gated by requiring every equation to have either (i) a \texttt{verified} stamp or (ii) a documented entry in \texttt{HUMAN\_REVIEW.md}. If a statement does get corrected, it will be logged in an ERRATA file, keeping track of the mistakes made by agents and allowing further error analysis.

\subsection{Prove} 

After the Extract and Verify stages, we have the full list of theorems that are faithful but await to be proved. An agent surveys all open \texttt{sorry}s and proposes a proof plan. It contains a methods component, consisting of general proof strategies that apply to many identities, and waves of proofs that resolve the remaining theorems in dependency order. This step also observes and modifies the DAG dependency structure. The proving stage is gated on either (i) a \texttt{sorry}-free proof body or (ii) documentation in \texttt{UNPROVEN.md}.

\subsection{Audit}

Beyond the pipeline, we have auditor agents to inspect the human review documents at each stage (dispositions, human\_review, unproven, etc.). Very often, these documents contain a mix of simple mistakes, such as a wrong convention, and serious issues, such as a hidden assumption that is not stated in the informal text. The auditors are prompted with the human evaluation guidelines and make simple fixes while providing analysis and suggestions for the harder issues that really need a human decision.

We would like to emphasize the need for gating and human interaction at each stage. Asking the agent to work alone risks two failure modes: the agent spends many tokens working on a single problem and gets stuck, or it secretly skips the work and does not report its true status. Human review is also crucial for design choices, such as definition conventions, or special cases in which implicit constraints or specifications are conveyed but not listed. 

Having the auditor significantly reduces the work of a human expert. The human expert can focus on the hard and subtle cases in which we do not readily trust the model's judgement, while the rest are taken care of autonomously. Having an auditor does not reduce the reliability of the output, since the edited statement will have to go through the verification check again before it can be stamped \texttt{verified}. On very special occasions, for example, one that involves a source misprint, a human can stamp a theorem \texttt{verified} despite it not passing the verification check, but that is only possible after careful review by the expert, and no autonomous agents can do that. 

Combined with our verification method, the workflow allows autoformalization of large-scale libraries without losing credibility. As part of this work in progress, we were able to complete the formalization of about 10\% of DLMF, comprising around 1,000 entries, in 2 man-months. Further details are available in Appendix~\ref{app:chapter-progress}.

\section{Experiments}
\label{sec:experiments}

In the following experiments, we compare our verification method with other baseline methods across different datasets. We demonstrate that ours has the lowest rate of accepting an incorrect formalization, which is the goal of building a reliable autoformalization verification. We also report runs of our method on external datasets, presenting potential defects in them. 

\subsection{Setup}
\label{sec:setup}

\paragraph{Datasets.} The experiment contains 1,810 informal--formal pairs from four datasets, and the datasets provide a correctness evaluation for each pair. \textbf{FLMF} is a selected set of 471 statements from the ongoing DLMF formalization, combining library statements, mutations\footnote{We included some mutations that are intended to be incorrect, mutation mechanisms are extracted from historical errors}, historical errors, and source-misprint pairs. This dataset focuses on special function equalities. \textbf{ProofNet} uses the original release \citep{azerbayev2023proofnet}, with Ground False's faithfulness verdicts as labels \citep{an2026}. It contains 371 undergraduate math problems from analysis, algebra, and topology. \textbf{Beyond Compilation} is a set of 400 graduate-level math problems with semantic evaluations \citep{zhang2026beyond}. \textbf{RLM25} extracts 617 informal--formal statement pairs from six research-level formalization projects across number theory, harmonic analysis, combinatorics, etc\footnote{RLM25 doesn't contain any evaluation component, so we assume all statements are labeled ``correct''.}\citep{poiroux2025reliable}. Appendix~\ref{app:dataset} gives a more detailed breakdown of dataset construction process.

\paragraph{Methods and models.} \emph{Ours} is the union of the three checks as discussed in Section~\ref{sec:verification}, where direct inspection uses GPT-5.5, and back-translation uses Gemini 3.1 Pro as the translator and GPT-5.5 as the judge. The baseline methods are GPT-5.5 and Opus 4.6 as direct judges, each called once per pair and once in joint batches of 16 (``joint''); LeanScorer \citep{xuejun2025mathesis}, which uses LLMs (GPT-5.5) to assess how faithfully each component of a statement is formalized; and FormalAlign \citep{lu2025formalalign}, a trained model that evaluates alignment between informal and formal statements, at a certainty threshold of 0.3 (Appendix~\ref{app:formalalign} gives more detail). Every method sees the same informal statement and formal context.

\paragraph{Method coverage.} The semantic checks, direct LLM judges, LeanScorer, and FormalAlign apply to every informal--formal pair. However, the goal-independent part of the vacuity check requires a completed proof and has no coverage on the original ProofNet or Beyond Compilation. In these cases, the vacuity check only looks for trivial proof methods. Similarly, the numerical check requires a statement to be numerically evaluable, which is not the case for, for example, infinite sums statements.

\paragraph{Metric.} We evaluate acceptance by a verification method against each dataset's labels, which are categorized into ``correct'' and ``incorrect''. Our primary metric is the \emph{accepted-error rate}: among the items a verification method accepts (does not flag), the fraction that are labeled incorrect. This metric aligns with our goal of ensuring that a Lean statement passing the gate is correct. We also report the percentage of total statements flagged.

\subsection{Main results}

\begin{table}[t]
\centering
\scriptsize
\setlength{\tabcolsep}{4pt}
\renewcommand{\arraystretch}{1.1}
\begin{tabular}{@{}l cc c cc c cc@{}}
\toprule
& \multicolumn{2}{c}{\shortstack{FLMF\\(182 C / 289 I)}} && \multicolumn{2}{c}{\shortstack{Beyond Compilation\\(245 C / 106 I)}} && \multicolumn{2}{c}{\shortstack{ProofNet\\(163 C / 204 I)}} \\
\cmidrule{2-3}\cmidrule{5-6}\cmidrule{8-9}
Method & \shortstack{Incorrect / accepted\\(rate)} & \shortstack{Flagged / \\ correct} && \shortstack{Incorrect / accepted\\(rate)} & \shortstack{Flagged / \\correct} && \shortstack{Incorrect / accepted\\(rate)} & \shortstack{Flagged / \\ correct} \\
\midrule
Ours & \textbf{5/105 (4.8\%)} & 45.1\% && \textbf{3/154 (1.9\%)} & 38.4\% && \textbf{14/153 (9.2\%)} & 14.7\% \\
GPT-5.5 & 28/178 (15.7\%) & 17.6\% && 10/227 (4.4\%) & 11.4\% && 26/176 (14.8\%) & 8.0\% \\
GPT-5.5 (joint) & 45/203 (22.2\%) & 13.2\% && 13/232 (5.6\%) & 10.6\% && 38/194 (19.6\%) & 4.3\% \\
Opus 4.6 & 72/232 (31.0\%) & 12.1\% && 24/261 (9.2\%) & 3.3\% && 44/206 (21.4\%) & 0.6\% \\
Opus 4.6 (joint) & 88/250 (35.2\%) & 11.0\% && 42/267 (15.7\%) & 8.2\% && 56/215 (26.0\%) & 2.5\% \\
LeanScorer & 10/93 (10.8\%) & 54.4\% && 7/175 (4.0\%) & 31.4\% && 17/135 (12.6\%) & 27.6\% \\
FormalAlign ($\le 0.3$) & 32/76 (42.1\%) & 75.8\% && 49/139 (35.3\%) & 63.3\% && 16/20 (80.0\%) & 97.5\% \\
\bottomrule
\end{tabular}
\caption{Accepted-error rate by method on the three labeled datasets. \emph{Incorrect / accepted} counts the incorrect items among those a method did not flag. A lower accepted-error rate means fewer incorrect Lean formalizations after the check. \emph{Flagged / correct} is the percentage of all items the method flags among those labeled ``correct''. The counts under each dataset are its correct (C) and incorrect (I) items.}
\label{tab:main}
\end{table}

Table~\ref{tab:main} reports the results of running all methods on the FLMF, Beyond Compilation, and ProofNet datasets. Our method's accepted-error rates are the lowest reported rates on all three datasets. Admittedly, our flagging rate is towards the higher end among all methods, but false alarms are generally acceptable given the help of auditing agents. We note that the ``incorrect'' labels from the datasets are not always accurate. The 3 items labeled incorrect that our method accepted in Beyond Compilation turn out to all have source ambiguity or incorrect labeling, and we would actually label 2 of the 3 ``correct'' after expert review. 

In Table~\ref{tab:flag-review}, we present an audit using our verification method on the Beyond Compilation, ProofNet, and RLM25 datasets. The flags raised by the verification pipeline then go through expert review to identify translation defects, source defects, or false alarms. In each of them, we were able to identify a number of defective statements that were labeled ``correct''. We will release the corrected dataset with this paper. 

Notice that the flags raised by a numerical check have a very high rate of identifying real defects, and many of those defects were not identified by a semantic LLM check. In Beyond Compilation, all 41 statements flagged by the numerical check yet labeled ``correct'' turn out to contain real defects. In RLM25, 8 out of the 12 flagged statements contain real defects. The four RLM25 numerical false alarms are attributed to binding errors in adapting the checker to the project's data structures.

\begin{table}
\centering
\scriptsize
\setlength{\tabcolsep}{4pt}
\renewcommand{\arraystretch}{1.05}
\begin{tabular}{@{}llrrrrr@{}}
\toprule
Dataset & Check & Flagged & \shortstack{Labeled\\incorrect} & \shortstack{Translation\\defects} & \shortstack{Source\\defects} & \shortstack{False\\alarm} \\
\midrule
Beyond Compilation & Numerical & 76 & 35 & 8 & 37 & 0 \\
 & Semantic & 151 & 101 & 15 & 12 & 25 \\
 & Vacuity & 23 & 12 & 1 & 0 & 10 \\
 & Combined & 195 & 102 & 18 & 43 & 35 \\
\addlinespace
ProofNet & Numerical & 85 & 85 & 0 & 0 & 0 \\
 & Semantic & 203 & 183 & 1 & 1 & 17 \\
 & Vacuity & 4 & 1 & 0 & 0 & 3 \\
 & Combined & 208 & 186 & 1 & 1 & 19 \\
\addlinespace
RLM25 & Numerical & 12 & n/a & 7 & 0 & 4 \\
 & Semantic & 367 & n/a & 184 & 21 & 143 \\
 & Vacuity & 12 & n/a & 3 & 1 & 8 \\
 & Combined & 373 & n/a & 184 & 21 & 149 \\
\bottomrule
\end{tabular}
\caption{Flags across different datasets and reviewed defect types. \emph{Flagged} counts all alerts by each method. \emph{Labeled incorrect} counts alerts on items the dataset labels incorrect. \emph{Translation defects} and \emph{Source defects} count review findings among flagged items that the dataset labels correct. RLM25 has no correctness labels; therefore, all statements are considered ``correct'' by the dataset.}
\label{tab:flag-review}
\end{table}

\section{Case Studies}
\label{sec:cases}

We present two case studies on formalization defects identified by our method. One is identified by the vacuity checker alone and not by any other method, which is a subtle formalization defect involving contradictory hypotheses. The second is a false Taylor bound from the \textit{Beyond Compilation} dataset that was faithfully translated to Lean and flagged only by our numerical check and none others. We therefore argue that even if we can make sure an LLM translates an informal statement seemingly well, the statement itself can contain subtleties or defects that are not captured by the model, resulting in an incorrect formalization.

\subsection{Case 1: vacuous statement in FLMF.} 

\paragraph{DLMF \S7.13(iii)}

In the first quadrant of $\mathbb{C}$, the Fresnel integral $C(z)$ has an infinite set of zeros $z_n=x_n+iy_n$, $n=1,2,3,...$ arranged in order of increasing absolute value. As $n\to\infty$, the $x_n$ and $y_n$ corresponding to the zeros of $C(z)$ satisfy 
\begin{equation}
x_n \sim \lambda + \frac{\alpha(\alpha\pi-4)}{8\pi\lambda^3}+\cdots,
\qquad
y_n \sim \frac{\alpha}{2\lambda}+\cdots,
\tag{DLMF \S7.13.5}
\end{equation}
with
\begin{equation}
\lambda=\sqrt{4n-1},
\qquad
\alpha=(2/\pi)\ln(\pi\lambda).
\tag{DLMF \S7.13.6}
\end{equation}

The extracted Lean statement was (definitions stripped)
\begin{quote}\scriptsize\ttfamily\raggedright
theorem erf\_7\_13\_5 (z : $\mathbb{N}\to\mathbb{C}$)\\
\hspace*{1em}(hzero : $\forall$ n, cfresnelC (z n) = 0)\\
\hspace*{1em}(hquad : $\forall$ n, 0 < (z n).re $\wedge$ 0 < (z n).im)\\
\hspace*{1em}(hmono : StrictMono (fun n $\mapsto$ (z n).re))\\
\hspace*{1em}(hall : $\forall$ w : $\mathbb{C}$, cfresnelC w = 0 $\to$\\
\hspace*{5em}0 < w.re $\to$ 0 < w.im $\to$ $\exists$ n, z n = w) :\\
\hspace*{1em}(fun n $\mapsto$ (z n).re - fresnelCZeroReApprox n)\\
\hspace*{2em}=o[atTop] (fun n $\mapsto$ fresnelCZeroLambda n \^{} ($-3$ : $\mathbb{Z}$)) $\wedge$\\
\hspace*{1em}(fun n $\mapsto$ (z n).im - fresnelCZeroImApprox n)\\
\hspace*{2em}=o[atTop] (fun n $\mapsto$ fresnelCZeroLambda n \^{} ($-1$ : $\mathbb{Z}$)) := by
\end{quote}

This Lean theorem assumes that a sequence $z_n$ enumerates all first-quadrant zeros of $C(z)$ while their \textbf{real parts} are strictly increasing (\texttt{hmono}). However, the informal source explicitly orders the zeros by increasing \textbf{absolute value}. The assumptions in the Lean statement are contradictory because the large-index zeros will approach the real axis, and a mirror symmetry transforms them into first-quadrant zeros with arbitrarily small positive real parts. Therefore, there must be a first-quadrant zero whose real part is smaller than that of $z_0$, the first entry of the sequence, violating the monotonicity imposed by \texttt{hmono}. The resulting contradiction among the hypotheses proves \texttt{False}, from which the stated asymptotic conclusion can be obtained vacuously by \texttt{False.elim}.

In verification, this Lean statement obtained a \texttt{FAITHFUL} verdict from the semantic checks because the checker was not able to catch this subtle mistake. Indeed, all other baseline methods accepted it as a correct translation, and the proof compiled without error. Only the vacuity check was able to spot the problem.

\subsection{Case 2: A defective informal source in Beyond Compilation.}
The informal statement says that if $f$ is holomorphic on a neighborhood of the closed unit disk and $f(0)\ne0$, then there exists a constant $C\in\mathbb{R}$ such that
\begin{equation*}
\left|1-\sum_{k=0}^{m}\frac{f^{(k)}(0)}{k!\,f(0)}z^k\right|
\le C|z|^{m+1}
\qquad(m\in\mathbb{N},\ |z|\le1).
\end{equation*}
The corresponding formalization, with its theorem name abbreviated to \texttt{X}, is
\begin{quote}\scriptsize\ttfamily\raggedright
theorem X \{f : $\mathbb{C}\to\mathbb{C}$\}\\
\hspace*{1em}(hf : $\exists$ U : Set $\mathbb{C}$, IsOpen U $\wedge$ Metric.closedBall (0 : $\mathbb{C}$) 1 $\subseteq$ U $\wedge$\\
\hspace*{2em}DifferentiableOn $\mathbb{C}$ f U)\\
\hspace*{1em}(h0 : f 0 $\ne$ 0) :\\
\hspace*{1em}$\exists$ C : $\mathbb{R}$, $\forall$ m : $\mathbb{N}$, $\forall$ z : $\mathbb{C}$, $\lVert$z$\rVert$ $\le$ 1 $\to$\\
\hspace*{2em}$\lVert$(1 : $\mathbb{C}$) - ($\sum$ k $\in$ Finset.range (m + 1),\\
\hspace*{4em}((deriv\^{}[k] f) 0) / ((k.factorial : $\mathbb{N}$) * f 0) * z \^{} k)$\rVert$\\
\hspace*{3em}$\le$ C * $\lVert$z$\rVert$ \^{} (m + 1)
\end{quote}

This is a faithful translation of the informal statement. However, the numerical check finds the admissible counter example $f(z)=e^z$. With $m=1$, for real $0<z\le1$ the claimed bound becomes
\begin{equation*}
|1-(1+z)|=|z|\le C|z|^2.
\end{equation*}
Dividing both sides by $|z|^2$, we get $C\ge1/|z|$, which no constant $C$ can satisfy as $z\to 0$. 

In the experiment, this correct-labeled entry is flagged only by the numerical check among all methods. Back-translation, GPT-5.5 and Opus 4.6 in both single-pair and joint modes return \texttt{FAITHFUL}; LeanScorer accepts it with a score of $0.67$ at a threshold of $0.6$; FormalAlign accepts it with a certainty score of $0.394$ at a threshold of $0.3$. None of the stored semantic explanations identifies the impossible uniform bound. 

\section{Conclusion}
\label{sec:conclusion}

In this paper, we introduced our verification method, conducted experiments across different benchmarks, and demonstrated that our method yields the lowest accepted-error rate, therefore yielding the most trustworthy outcome. We integrate these methods into an Extract--Verify--Prove pipeline with human gates and audits,  ensuring completeness and faithfulness of the formalization results. FLMF provides the development setting and a preliminary evaluation panel, and its broader dataset remains under construction and will be presented in a separate paper.

We highlight the numerical and vacuity checks in our verification, which are complementary to the verdicts of an LLM or model-based judge, the most common approach used across all other formalization work. According to our experiments, these checks are able to spot mistakes ignored by all LLM judges, including subtle translation defects or informal source errors. The checks are also informative in pointing out the exact numerical misalignment or proof inconsistency. A theorem passing these checks can be trusted to be at least numerically consistent and not trivially vacuous. Combined with our autoformalization pipeline, we can then obtain a promising Lean formalization of the informal source.

There are certain limitations to our current methods. Numerical and vacuity checks are only applicable to numerically evaluable theorems or complete proofs. There could be defects in these checks as well, for example, a wrong binding in the numerical check or a subtly vacuous proof that does not fall within the detection range of our vacuity detector. Passing these checks doesn't guarantee correctness either, for example a narrower domain in the Lean statement compared to the informal source is a common failure mode. 

Future work will extend the numerical and vacuity checks to handle additional corner cases encountered during formalization, while guaranteeing faithfulness of the checks. We aim to reduce vacuity false alarms by distinguishing contradictory assumptions and missing mathematical content from legitimate definitional identities and routine proofs. We will also explore additional mechanical checks that complement the existing methods and provide independently checkable evidence. 

\section*{Acknowledgements}

We thank Shi Chen, Yongxi Lin, Leyan Pan, Fred Rajasekaren, Christian Szegedy, Zixiao Wang, and Ruofan Zeng for valuable discussions. Most of this work was completed during AM's internship at AletheAI in summer 2026, and we are grateful to AletheAI for the resources and support that made this work possible. We also thank the DLMF committee at NIST for their valuable feedback.

{\small\raggedright

}
\clearpage
\appendix
\section{Chapter-level pipeline progress}
\label{app:chapter-progress}

Table~\ref{tab:chapter-progress} reports progress of FLMF on all 36 chapters. Of the 8,833 numbered DLMF equations, 1,326 (15\%) are stated in FLMF, 641 (7\%) are verified, and 1,142 have a \texttt{sorry}-free proof. Chapter 5, 7, and 20 are essentially complete. Chapters 6 and 15 are fully extracted but less fully verified (56 of 97 and 52 of 198).

\begingroup
\scriptsize
\setlength{\tabcolsep}{3pt}
\renewcommand{\arraystretch}{1.04}
\begin{longtable}{@{}rL{.42\linewidth}rrrr@{}}
\caption{Complete DLMF chapter progress, reported as counts.}
\label{tab:chapter-progress}\\
\toprule
Ch. & Chapter & Targets & Extract & Verified & Proved \\
\midrule
\endfirsthead
\multicolumn{6}{l}{\textit{Table~\ref{tab:chapter-progress} continued}}\\
\toprule
Ch. & Chapter & Targets & Extract & Verified & Proved \\
\midrule
\endhead
\midrule
\multicolumn{6}{r}{\textit{continued on next page}}\\
\endfoot
\bottomrule
\endlastfoot
1 & Algebraic and Analytic Methods & 705 & 88 & 38 & 75 \\
2 & Asymptotic Approximations & 349 & 2 & 1 & 2 \\
3 & Numerical Methods & 297 & 2 & 0 & 2 \\
4 & Elementary Functions & 518 & 26 & 14 & 24 \\
5 & Gamma Function & 164 & 162 & 123 & 137 \\
6 & Exponential, Logarithmic, Sine and Cosine Integrals & 100 & 97 & 56 & 77 \\
7 & Error Functions, Dawson's and Fresnel Integrals & 137 & 137 & 136 & 137 \\
8 & Incomplete Gamma and Related Functions & 241 & 15 & 1 & 15 \\
9 & Airy and Related Functions & 230 & 13 & 4 & 11 \\
10 & Bessel Functions & 698 & 52 & 10 & 40 \\
11 & Struve and Related Functions & 149 & 11 & 6 & 11 \\
12 & Parabolic Cylinder Functions & 177 & 4 & 1 & 4 \\
13 & Confluent Hypergeometric Functions & 381 & 17 & 3 & 17 \\
14 & Legendre and Related Functions & 284 & 14 & 4 & 11 \\
15 & Hypergeometric Function ${}_2F_1$ & 202 & 198 & 52 & 157 \\
16 & Generalized Hypergeometric and Meijer $G$ & 99 & 5 & 2 & 5 \\
17 & $q$-Hypergeometric and Related Functions & 182 & 7 & 1 & 7 \\
18 & Orthogonal Polynomials & 677 & 56 & 10 & 43 \\
19 & Elliptic Integrals & 464 & 17 & 3 & 17 \\
20 & Theta Functions & 113 & 113 & 88 & 112 \\
21 & Multidimensional Theta Functions & 59 & 0 & n/a & n/a \\
22 & Jacobian Elliptic Functions & 258 & 22 & 10 & 18 \\
23 & Weierstrass Elliptic and Modular Functions & 167 & 21 & 6 & 21 \\
24 & Bernoulli and Euler Polynomials & 188 & 22 & 7 & 20 \\
25 & Zeta and Related Functions & 174 & 24 & 10 & 23 \\
26 & Combinatorial Analysis & 201 & 16 & 9 & 16 \\
27 & Functions of Number Theory & 132 & 33 & 20 & 29 \\
28 & Mathieu Functions and Hill's Equation & 353 & 15 & 1 & 15 \\
29 & Lam\'e Functions & 205 & 1 & 0 & 1 \\
30 & Spheroidal Wave Functions & 114 & 3 & 0 & 2 \\
31 & Heun Functions & 120 & 103 & 25 & 64 \\
32 & Painlev\'e Transcendents & 286 & 6 & 0 & 6 \\
33 & Coulomb Functions & 146 & 0 & n/a & n/a \\
34 & $3j$, $6j$, $9j$ Symbols & 64 & 18 & 0 & 17 \\
35 & Functions of Matrix Argument & 62 & 0 & n/a & n/a \\
36 & Integrals with Coalescing Saddles & 137 & 6 & 0 & 6 \\
\midrule
& \textbf{Total} & \textbf{8,833} & \textbf{1,326} & \textbf{641} & \textbf{1,142} \\
\end{longtable}
\endgroup

\clearpage
\section{DLMF-side discrepancies identified to date}
\label{app:dlmf-misprints}

The following potential misprints in the DLMF had been identified while formalizing and proving the corresponding FLMF statements. They have been confirmed by the DLMF committee and corrections are updated.

\begin{table}[H]
\centering
\small
\setlength{\tabcolsep}{4pt}
\begin{tabular}{@{}L{.13\linewidth}L{.36\linewidth}L{.45\linewidth}@{}}
\toprule
DLMF & Printed defect and correction & Independent evidence \\
\midrule
20.10.4 & The display uses $(1+\beta)$ where scale covariance and the theta shift require $(\ell+\beta)$. The $\theta_1$ transform has the opposite printed sign, and the $\theta_1$ and shifted $\theta_2$ integrals are negatives of each other, not equal. & Theta quasi-periodicity already makes the printed chain inconsistent at $\ell=1$; termwise Laplace transforms and independent 25-digit quadrature recover the corrected form. \\
\addlinespace
20.10.5 & The display uses $(1+\beta)$ where scale covariance and the theta shift require $(\ell+\beta)$. & Termwise Laplace transforms and independent 25-digit quadrature recover the corrected form. \\
\addlinespace
20.13.5 & The alternating heat-kernel sum is printed as $\theta_4$; Poisson summation gives $\theta_2$. & At $(\alpha,t,z)=(0.7,0.3,0.4)$, the sum is $1.5982\ldots$ and matches $\theta_2$ to $1.6\times10^{-26}$, whereas the printed $\theta_4$ side is $0.3972\ldots$; complex samples agree. \\
\bottomrule
\end{tabular}
\caption{DLMF misprint findings.}
\label{tab:dlmf-misprints}
\end{table}

\clearpage
\section{DLMF extraction example}
\label{app:extraction}

The following is the presentation MathML (pMML) source for DLMF \S 5.5.3. 

\begin{Verbatim}[fontsize=\scriptsize,baselinestretch=0.92,breaklines=true,breakanywhere=true]
<m:math xmlns:m="http://www.w3.org/1998/Math/MathML"
  altimg="E3.png" altimg-height="99px" altimg-valign="-7px"
  altimg-width="149px"
  alttext="\Gamma\left(z\right)\Gamma\left(1-z\right)=\pi/\sin\left(\pi z\right),"
  display="block">
  <m:mrow>
    <m:mrow>
      <m:mrow>
        <m:mrow>
          <m:mi href="DLMF:/5.2#E1" mathvariant="normal"
            title="gamma function">&#x393;</m:mi>
          <m:mo>&#x2061;</m:mo>
          <m:mrow><m:mo>(</m:mo>
            <m:mi href="DLMF:/5.1#t1.r4"
              title="complex variable">z</m:mi>
            <m:mo>)</m:mo></m:mrow>
        </m:mrow>
        <m:mo>&#x2062;</m:mo>
        <m:mrow>
          <m:mi href="DLMF:/5.2#E1" mathvariant="normal"
            title="gamma function">&#x393;</m:mi>
          <m:mo>&#x2061;</m:mo>
          <m:mrow><m:mo>(</m:mo><m:mrow><m:mn>1</m:mn>
            <m:mo>&#x2212;</m:mo>
            <m:mi href="DLMF:/5.1#t1.r4"
              title="complex variable">z</m:mi>
          </m:mrow><m:mo>)</m:mo></m:mrow>
        </m:mrow>
      </m:mrow>
      <m:mo>=</m:mo>
      <m:mrow>
        <m:mi href="DLMF:/3.12#E1"
          title="the ratio of the circumference of a circle to its diameter">
          &#x3C0;</m:mi>
        <m:mo>/</m:mo>
        <m:mrow>
          <m:mi href="DLMF:/4.14#E1" title="sine function">sin</m:mi>
          <m:mo>&#x2061;</m:mo>
          <m:mrow><m:mo>(</m:mo><m:mrow>
            <m:mi href="DLMF:/3.12#E1"
              title="the ratio of the circumference of a circle to its diameter">
              &#x3C0;</m:mi>
            <m:mo>&#x2062;</m:mo>
            <m:mi href="DLMF:/5.1#t1.r4"
              title="complex variable">z</m:mi>
          </m:mrow><m:mo>)</m:mo></m:mrow>
        </m:mrow>
      </m:mrow>
    </m:mrow>
    <m:mo>,</m:mo>
  </m:mrow>
</m:math>
\end{Verbatim}

Because the standalone pMML omits the side condition, the extractor reads the
equation's container in the full DLMF section page. Its normalized JSON output is:

\begin{Verbatim}[fontsize=\scriptsize,baselinestretch=0.92,breaklines=true,breakanywhere=true]
{
 "tag": "5.5.3",
 "url": "https://dlmf.nist.gov/5.5#E3",
 "kind": "equation",
 "latex": [
  "\\Gamma\\left(z\\right)\\Gamma\\left(1-z\\right)=\\pi/\\sin\\left(\\pi z\\right),",
  "\\Gamma\\left(\\NVar{z}\\right)",
  "z\\neq 0,\\pm 1,\\dots",
  "\\sin\\NVar{z}",
  "\\pi",
  "z"
 ],
 "eqs": [],
 "section_eqs": [
  [
   "5.5.1",
   "\\Gamma\\left(z+1\\right)=z\\Gamma\\left(z\\right),"
  ],
  [
   "5.5.2",
   "\\psi\\left(z+1\\right)=\\psi\\left(z\\right)+\\frac{1}{z}."
  ],
  [
   "5.5.3",
   "\\Gamma\\left(z\\right)\\Gamma\\left(1-z\\right)=\\pi/\\sin\\left(\\pi z\\right),"
  ],
  [
   "5.5.4",
   "\\psi\\left(z\\right)-\\psi\\left(1-z\\right)=-\\pi/\\tan\\left(\\pi z\\right),"
  ],
  [
   "5.5.5",
   "\\Gamma\\left(2z\\right)=\\pi^{-1/2}2^{2z-1}\\Gamma\\left(z\\right)\\Gamma\\left(z+\\tfrac{1}{2}\\right)."
  ],
  [
   "5.5.6",
   "\\Gamma\\left(nz\\right)=(2\\pi)^{(1-n)/2}n^{nz-(1/2)}\\prod_{k=0}^{n-1}\\Gamma\\left(z+\\frac{k}{n}\\right)."
  ],
  [
   "5.5.7",
   "\\prod_{k=1}^{n-1}\\Gamma\\left(\\frac{k}{n}\\right)=(2\\pi)^{(n-1)/2}n^{-1/2}."
  ],
  [
   "5.5.8",
   "\\psi\\left(2z\\right)=\\tfrac{1}{2}\\left(\\psi\\left(z\\right)+\\psi\\left(z+\\tfrac{1}{2}\\right)\\right)+\\ln 2,"
  ],
  [
   "5.5.9",
   "\\psi\\left(nz\\right)=\\frac{1}{n}\\sum_{k=0}^{n-1}\\psi\\left(z+\\frac{k}{n}\\right)+\\ln n."
  ]
 ],
 "context": ", pMML , png See also: Annotations for \u00a75.5(i) , \u00a75.5 and Ch.5 \u00a75.5(ii) Reflection \u24d8 Keywords: gamma function , psi function , reflection formula Notes: See Olver ( 1997b , pp. 35 and 39) . Permalink: http://dlmf.nist.gov/5.5.ii See also: Annotations for \u00a75.5 and Ch.5 5.5.3 \u0393 \u2061 ( z ) \u2062 \u0393 \u2061 ( 1 \u2212 z ) = \u03c0 / sin \u2061 ( \u03c0 \u2062 z ) , z \u2260 0 , \u00b1 1 , \u2026 , \u24d8 Symbols: \u0393 \u2061 ( z ) : gamma function , \u03c0 : the ratio of the circumference of a circle to its diameter , sin \u2061 z : sine function and z : complex variable A&S Ref: 6.1.17 (without the condition on z .) Referenced by: \u00a710.19(i) , \u00a715.4(ii) , \u00a715.8(ii) , (25.9.2) , \u00a75.21 , (9.10.18) , (9.12.15) Permalink: http://dlmf.nist.gov/5.5.E3 Encodings: TeX , pMML , png See also: Annotations for \u00a75.5(ii) , \u00a75.5 and Ch.5 <table",
 "xref_context": [],
 "constraints": [
  "z\\neq 0,\\pm 1,\\dots"
 ],
 "symbols": [
  {
   "token": "\u0393",
   "href": "5.2#E1",
   "gloss": "gamma function"
  },
  {
   "token": "z",
   "href": "5.1#t1.r4",
   "gloss": "complex variable"
  },
  {
   "token": "\u03c0",
   "href": "3.12#E1",
   "gloss": "the ratio of the circumference of a circle to its diameter"
  },
  {
   "token": "sin",
   "href": "4.14#E1",
   "gloss": "sine function"
  }
 ],
 "info": {
  "referenced_by": "\u00a710.19(i) , \u00a715.4(ii) , \u00a715.8(ii) , (25.9.2) , \u00a75.21 , (9.10.18) , (9.12.15)",
  "as_ref": "6.1.17 (without the condition on z .)",
  "proof_sketch": "",
  "sources": "",
  "keywords": "",
  "notes": "",
  "defines": "",
  "symbols_text": "\\Gamma\\left(\\NVar{z}\\right) : gamma function , \\pi : the ratio of the circumference of a circle to its diameter , \\sin\\NVar{z} : sine function and z : complex variable",
  "changes": []
 },
 "parser_version": 2
}
\end{Verbatim}

\clearpage
\section{Dataset construction, panel composition, and detection}
\label{app:dataset}
\raggedbottom

\subsection{Dataset construction and labels}

The evaluation contains 1,810 informal--formal pairs from four datasets. Table~\ref{tab:panel} gives their composition. Each dataset supplies its own criteria for labeling a pair correct or incorrect; these labels do not represent a common judgment of faithfulness or mathematical validity.

\paragraph{FLMF (182 correct / 289 incorrect).} The FLMF dataset contains 471 statements in eight components: 40 library seeds, 40 copies with bound variables renamed, 7 human-approved corrections to DLMF misprints, 160 intentional mutations of the seeds, 17 historical FLMF errors, 7 printed forms of DLMF misprints, 100 sampled verified theorems, and 100 statements from the project's errata log. The seeds and renamings each contain 38 correct and 2 incorrect items, and the sampled verified theorems contain 99 correct and 1 incorrect item. Corrected misprints are labeled correct. Mutations, historical errors, printed misprints, and errata items are labeled incorrect.

The mutations target eight mechanisms (M1--M8): impossible hypotheses, wrong objects, branches, totalization at poles, quantifier strength, indices and coercions, normalization, and lost source content. There are 12--20 mutations per mechanism and 24 simpler sanity mutations (Table~\ref{tab:mechanism}). We exclude 20 independent control nominations. We also exclude two Fresnel coefficient pairs, comprising four items, because their stronger asymptotic claims are unresolved. Historical Fresnel errors remain included.

The 100 sampled verified theorems satisfy the same eligibility criteria as the seeds: a semantic \texttt{verified} stamp at the pinned commit, a single numbered DLMF equation, and a completed proof. Selection is deterministic and stratified by library area, with a cap of 20 items per stratum. This component is disjoint from the seeds, renamings, mutations, historical artifacts, and misprint pairs by both declaration name and DLMF equation.

Each of the 100 errata items is a declaration recovered from git history in its form immediately before correction. Of these, 76 correspond to errors found by the numerical or vacuity check and 24 to errors found by other methods. They form 65 erratum clusters, and 99 of the 100 statements elaborate. Together with the mutations, historical errors, printed misprints, and five library entries rejected in human review, they account for the 289 incorrect items. The 129 incorrect items other than mutations provide examples of errors arising in formalization or in the source.

\paragraph{ProofNet (163 correct / 204 incorrect)} \citep{azerbayev2023proofnet}. We use 367 original ProofNet statements from the Lean 4.28 port distributed with ProofNet-Verified. The formal statements are those of ProofNet, and the labels are the Ground False verdicts \citep{an2026}. The 163 items labeled \emph{faithful} are correct. The incorrect items comprise 75 labeled \emph{stronger}, 55 \emph{weaker}, 64 \emph{incomparable}, and 10 whose informal statement is ambiguous (9) or wrong (1).

\paragraph{Beyond Compilation (245 correct / 106 incorrect)} \citep{zhang2026beyond}. We use 351 of the 400 released sources. The exclusions are 42 reported compilation failures, 6 entries from which no single statement can be separated from the proof, and 1 informal source that states no claim. A pair is labeled correct if the release's GPT-5.2 judge assigns a score of at least 9. Human annotations are available for 156 items whose recorded judge score matches the candidate's released score, and we use them as secondary labels. The annotations are predominantly acceptances and concern translations selected from model acceptances, so they do not independently assess the items labeled incorrect by the judge.

\paragraph{RLM25 (621 reference pairs)} \citep{poiroux2025reliable}. The reconstructed dataset contains reference pairs from six projects: Carleson (111), FLT (56), FLT3 (84), PFR (146), PrimeNumberTheoremAnd (99), and TestingLowerBounds (125). The published release contains 619 pairs. Reconstruction from public sources yields 621, including one extra eligible node in PFR and one in TestingLowerBounds, so membership does not exactly match the release manifest. RLM25 provides no correctness labels. We treat all reference pairs as correct and report their flag rate.

The formal context includes project definitions referenced by each statement, including definitions declared in other files and those introduced with \texttt{irreducible\_def} or \texttt{noncomputable abbrev}. For 310 of the 621 items, this supplies definitions absent from the source prefix in the statement's own file. Statement extraction accounts for local \texttt{let} expressions, which is necessary to preserve two statements in full. FormalAlign's released adapter does not receive formal context.

\paragraph{Proof availability.} The goal-independent part of the vacuity check requires a completed proof. The 100 sampled verified FLMF theorems have completed proofs. For errata items, coverage depends on the availability of a replayable proof of the statement as written before correction. Beyond Compilation and the original ProofNet statements provide no proofs, so the vacuity check on these datasets is limited to trivial proof methods, as described in Section~\ref{sec:setup}.

\subsection{Panel composition and detection}

Table~\ref{tab:panel} gives the composition of the four datasets. Table~\ref{tab:mechanism} reports flags on the 271 FLMF seeds, renamings, mutations, historical artifacts, and misprint items. 

\begin{table}[H]
\centering
\small
\setlength{\tabcolsep}{4pt}
\begin{tabular}{@{}L{.18\linewidth}L{.46\linewidth}rr@{}}
\toprule
\textbf{Dataset} & \textbf{Component} & \textbf{Correct} & \textbf{Incorrect} \\
\midrule
\textbf{FLMF} & Seeds & 38 & 2 \\
 & Bound-variable renamings of the seeds & 38 & 2 \\
 & Human-approved corrected forms of DLMF misprints & 7 & \\
 & Mutations of the seeds (M1--M8 + sanity) & & 160 \\
 & Historical FLMF errors & & 17 \\
 & Misprinted DLMF equations & & 7 \\
 & Sampled verified theorems after human review & 99 & 1 \\
 & Errata statements & & 100 \\
\cmidrule(l){2-4}
 & \textbf{FLMF total (471 items)} & \textbf{182} & \textbf{289} \\
\midrule
\multirow[t]{5}{=}{\textbf{ProofNet}\newline{\footnotesize (verdict by\newline Ground False)}}
 & faithful & 163 & \\
 & stronger & & 75 \\
 & weaker & & 55 \\
 & incomparable & & 64 \\
 & informal source ambiguous (9) or wrong (1) & & 10 \\
\midrule
\textbf{Beyond Compilation} & Full release (label: released judge score $\ge 9$) & 245 & 106 \\
\midrule
\textbf{RLM25} & Reconstructed reference pairs from all six projects, with project definitions & 621 & \\
\midrule
\textbf{Total} & & \textbf{1,211} & \textbf{599} \\
\bottomrule
\end{tabular}
\caption{Dataset composition (1,810 items).}
\label{tab:panel}
\end{table}

\begin{table}[H]
\centering
\scriptsize
\setlength{\tabcolsep}{4pt}
\begin{tabular}{@{}lr ccccccccc@{}}
\toprule
FLMF subset & $n$ & Ours & GPT-5.5 & \shortstack{GPT-5.5\\(joint)} & Opus 4.6 & LeanScorer & FormalAlign & N & B & V \\
\midrule
\multicolumn{11}{@{}l}{\emph{Seeds, renamings, and corrected misprints}}\\
Seeds & 40 & 12 & 8 & 4 & 3 & 22 & 37 & 0 & 9 & 3 \\
Renamed seeds & 40 & 15 & 7 & 4 & 4 & 21 & 36 & 0 & 12 & 3 \\
Corrected misprints & 7 & 7 & 7 & 7 & 6 & 3 & 1 & 0 & 7 & 0 \\
\midrule
\multicolumn{11}{@{}l}{\emph{Mutations, historical artifacts, and printed misprints}}\\
M1 impossible hypotheses & 20 & 20 & 20 & 20 & 19 & 20 & 18 & 0 & 20 & 0 \\
M2 wrong object & 16 & 16 & 16 & 16 & 15 & 16 & 16 & 15 & 16 & 0 \\
M3 branches/phases & 20 & 20 & 20 & 20 & 20 & 20 & 19 & 18 & 20 & 0 \\
M4 totalization/poles & 20 & 18 & 18 & 18 & 13 & 19 & 20 & 13 & 18 & 1 \\
M5 quantifiers/asymptotics & 16 & 16 & 14 & 14 & 14 & 16 & 14 & 9 & 15 & 0 \\
M6 indices/coercions & 16 & 16 & 16 & 16 & 16 & 16 & 16 & 14 & 16 & 0 \\
M7 normalization/scaling & 16 & 16 & 15 & 16 & 15 & 16 & 16 & 11 & 16 & 0 \\
M8 lost source content & 12 & 11 & 12 & 12 & 12 & 12 & 12 & 10 & 7 & 0 \\
Sanity mutations & 24 & 24 & 24 & 24 & 24 & 24 & 23 & 21 & 24 & 0 \\
Historical artifacts & 17 & 15 & 15 & 14 & 10 & 14 & 12 & 6 & 13 & 1 \\
Printed misprints & 7 & 7 & 6 & 2 & 4 & 4 & 1 & 5 & 6 & 0 \\
\bottomrule
\end{tabular}
\caption{Flags on the 271 FLMF seeds, renamings, mutations, historical artifacts, and misprint items. }
\label{tab:mechanism}
\end{table}

\section{FormalAlign threshold}
\label{app:formalalign}

FormalAlign returns a certainty score, and we flag a statement when its score is at most the chosen threshold. Table~\ref{tab:fa-sweep} reports the threshold sweep on the three labeled datasets. At the comparison threshold of 0.3, the accepted-error rates are 42.1\% on FLMF, 35.3\% on Beyond Compilation, and 80.0\% on ProofNet. Raising the threshold reduces the number of accepted statements but does not consistently reduce the fraction labeled incorrect. 

\begin{table}[H]
\centering
\small
\setlength{\tabcolsep}{4pt}
\renewcommand{\arraystretch}{1.15}
\begin{tabular}{@{}r cc cc cc@{}}
\toprule
& \multicolumn{2}{c}{\shortstack{FLMF\\(182 C / 289 I)}} & \multicolumn{2}{c}{\shortstack{Beyond Compilation\\(245 C / 106 I)}} & \multicolumn{2}{c}{\shortstack{ProofNet\\(163 C / 204 I)}} \\
\cmidrule(lr){2-3}\cmidrule(lr){4-5}\cmidrule(lr){6-7}
Threshold & \shortstack{Incorrect /\\accepted} & \shortstack{Flagged\\ / correct} & \shortstack{Incorrect /\\accepted} & \shortstack{Flagged\\ / correct} & \shortstack{Incorrect /\\accepted} & \shortstack{Flagged\\ / correct} \\
\midrule
0.1 & 243 / 409 & 8.8\% & 100 / 338 & 2.9\% & 185 / 312 & 22.1\% \\
0.2 & 111 / 199 & 51.6\% & 76 / 248 & 29.8\% & 81 / 120 & 76.1\% \\
\textbf{0.3} & \textbf{32 / 76} & \textbf{75.8\%} & \textbf{49 / 139} & \textbf{63.3\%} & \textbf{16 / 20} & \textbf{97.5\%} \\
0.4 & 12 / 22 & 94.5\% & 21 / 45 & 90.2\% & 2 / 2 & 100.0\% \\
0.5 & 1 / 1 & 100.0\% & 3 / 8 & 98.0\% & 0 / 0 & 100.0\% \\
\bottomrule
\end{tabular}
\caption{FormalAlign certainty-threshold sweep on 1,189 labeled pairs. A statement is flagged when its certainty score is at most the threshold. C and I denote the dataset's correct and incorrect labels. \emph{Incorrect / accepted} counts incorrect-labeled items among accepted statements. \emph{Flagged / correct} is the percentage of correct-labeled items that are flagged. The threshold of 0.3 used in the main comparison is shown in bold. The 0 / 0 entry indicates that no statements are accepted. RLM25 reference pairs are not included in this sweep.}
\label{tab:fa-sweep}
\end{table}

\section{Cost and runtime}
\label{app:cost}

\begin{table}[H]
\centering
\small
\setlength{\tabcolsep}{4pt}
\renewcommand{\arraystretch}{1.15}
\begin{tabular}{@{}lrrrrr@{}}
\toprule
Channel & Items & API calls & \shortstack{Cost\\(USD)}
& \shortstack{Cost / item\\(USD)} & \shortstack{Time / item\\(s)} \\
\midrule
Numerical check & 1,539 & 3,961 & 314.18 & 0.204 & 77.0 \\
Back-translation & 1,539 & 3,080 & 99.57 & 0.065 & 29.0 \\
Vacuity check & 1,539 & 0 & 0.00 & 0.000 & -- \\
GPT-5.5 direct & 918 & 920 & 33.01 & 0.036 & 13.5 \\
Opus 4.6 direct & 918 & 922 & 20.96 & 0.023 & 14.6 \\
LeanScorer & 918 & 1,840 & 66.65 & 0.073 & 20.5 \\
GPT-5.5 joint & 918 & 61 & 12.02 & 0.013 & 4.2 \\
Opus 4.6 joint & 918 & 61 & 5.43 & 0.006 & 3.1 \\
\bottomrule
\end{tabular}
\caption{API cost and recorded runtime.
The 918-item set comprises 200 FLMF, 351 Beyond Compilation,
and 367 ProofNet statements. The 1,539-item set also includes
621 RLM25 reference pairs.
Costs include retained reservations
totaling \$9.64 for 25 requests whose charges remain unresolved.
Time per item is the sum of final recorded job durations divided
by the item count, with joint-batch time amortized across items;
it excludes superseded attempts and is not elapsed wall-clock time.
Vacuity runtime is omitted because 1,512 records report result-reuse
overhead rather than the original checking time.
FormalAlign scored 918 items locally on CPU, with 32.1 minutes
of recorded inference time (2.1 seconds per item), excluding model loading.}
\label{tab:cost}
\end{table}

\end{document}